\documentclass[runningheads]{llncs}
\usepackage{graphicx}
\usepackage{bm}
\usepackage{amsfonts}
\usepackage{amsmath}
\usepackage{cleveref}
\usepackage{multirow}
\usepackage{booktabs}
\usepackage{algorithmic}
\usepackage{algorithm}
\usepackage{paralist}

\usepackage[dvipsnames]{xcolor}

\DeclareMathOperator{\sgn}{sgn}
\newcommand{\Loss}{\mathcal{L}}
\definecolor{light-gray}{gray}{0.6}

\usepackage{array}
\newcolumntype{P}[1]{>{\centering\arraybackslash}p{#1}}
\newcolumntype{R}[1]{>{\raggedright\arraybackslash}p{#1}}

\begin{document}
\newcommand*\samethanks[1][\value{footnote}]{\footnotemark[#1]}
%
\title{Audio Adversarial Examples for Robust Hybrid CTC/Attention Speech Recognition}
%
\titlerunning{Hybrid CTC/Attention Audio Adversarial Examples}
%
\author{Ludwig K{\"u}rzinger\orcidID{0000-0001-5312-3870}\thanks{These authors contributed equally to this work.} \and 
Edgar Ricardo Chavez Rosas\samethanks \and
Lujun Li\orcidID{0000-0002-0641-3178} \and
Tobias Watzel\orcidID{0000-0002-3552-3325} \and
Gerhard Rigoll\orcidID{0000-0003-1096-1596}}
\authorrunning{L. K{\"u}rzinger et al.}
\institute{Chair of Human-Machine Communication, Technical University of Munich, Germany\\
\email{\{ludwig.kuerzinger,ricardo.chavez\}@tum.de}}
\maketitle

\begin{abstract}
Recent advances in Automatic Speech Recognition (ASR) demonstrated how end-to-end systems are able to achieve state-of-the-art performance.
There is a trend towards deeper neural networks, however those ASR models are also more complex and prone against specially crafted noisy data.
Those Audio Adversarial Examples (AAE) were previously demonstrated on ASR systems that use Connectionist Temporal Classification (CTC), as well as attention-based encoder-decoder architectures.

Following the idea of the hybrid CTC/attention ASR system, this work proposes algorithms to generate AAEs to combine both approaches into a joint CTC-attention gradient method.
Evaluation is performed using a hybrid CTC/attention end-to-end ASR model on two reference sentences as case study,
as well as the TEDlium v2 speech recognition task.
We then demonstrate the application of this algorithm for adversarial training to obtain a more robust ASR model.

\end{abstract}

\section{Introduction}
In recent years, advances in GPU technology and machine learning libraries enabled the trend towards deeper neural networks in Automatic Speech Recognition (ASR) systems.
End-to-end ASR systems transcribe speech features to letters or tokens without any intermediate representations.
There are two major techniques:
\begin{inparaenum}[1)] 
    \item Connectionist Temporal Classification (CTC~\cite{graves2006connectionist}) carries the concept of hidden Markov states over to end-to-end neural networks as training loss for sequence classification networks.
    Neural networks trained with CTC loss calculate the posterior probability of each letter at a given time step in the input sequence.
    \item Attention-based encoder-decoder architectures such as~\cite{chan2016listen}, are trained as auto-regressive sequence-generative models. The encoder transforms the input sequence into a latent representation; from this, the decoder generates the sentence transcription.
\end{inparaenum}
The hybrid CTC/attention architecture combines these two approaches in one single neural network~\cite{watanabe2017hybrid}.

Our work is motivated by the observation that adding a small amount of specially crafted noise to a sample given to a neural network can cause the neural network to wrongly classify its input~\cite{szegedy2013intriguing}.
From the standpoint of system security, those algorithms have implications on possible attack scenarios. 
A news program or sound that was augmented with a barely noticeable noise can give hidden voice commands, e.g. to open the door, to the ASR system of a personal assistant~\cite{carlini2016hidden,carlini2018audio}.
From the perspective of ASR research, a network should be robust against such small perturbations that can change the transcription of an utterance;
its speech recognition capability shall relate more closely to what humans understand.

In speech recognition domain, working Audio Adversarial Examples (AAEs) were already demonstrated for CTC-based~\cite{carlini2018audio}, as well as for attention-based ASR systems~\cite{sun2019adversarial}.
The contribution of this work is a method for generation of untargeted adversarial examples in feature domain for the hybrid CTC/attention ASR system.
For this, we propose two novel algorithms that can be used to generate AAE for attention-based encoder-decoder architectures.
We then combine these with CTC-based AAEs to introduce an algorithm for joint CTC/attention AAE generation.
To further evaluate our methods and exploit the information within AAEs, the ASR network training is then augmented with generated AAEs.
Results indicate improved robustness of the model against adversarial examples, as well as a generally improved speech recognition performance by a moderate $10\%$ relative to the baseline model.


\section{Related Work}

\paragraph{Automatic Speech Recognition (ASR) Architecture.}
Our work builds on the hybrid CTC/attention ASR architecture as proposed and described in~\cite{watanabe2018espnet,watanabe2017hybrid}, using the location-aware attention mechanism~\cite{ChorowskiEtAl15}.
This framework combines the most two popular techniques in end-to-end ASR:
Connectionist Temporal Classification (CTC), as proposed in~\cite{graves2006connectionist}, and attention-based encoder-decoder architectures.
Attention-based sequence transcription was proposed in the field of machine language translation in~\cite{BahdanauEtAl14} and later applied to speech recognition in Listen-Attend-Spell~\cite{chan2016listen}.
Sentence transcription is performed with the help of a RNN language model (RNNLM) integrated into decoding process using {shallow fusion}~\cite{GulcehreEtAl15}.

\paragraph{Audio Adversarial Examples (AAEs).}
Adversarial examples were originally porposed and developed in the image recognition field and since then, they have been amply investigated in~\cite{szegedy2013intriguing,kurakin2016adversarialphysical,kurakin2016adversarial}.
The most known method for generation is the Fast Gradient Sign Method (FGSM)~\cite{goodfellow2014explaining}. 
Adversarial examples can be prompt to label leaking~\cite{kurakin2016adversarial}, that is when the model does not have difficulties finding the original class of the disguised sample, as the transformation from the original is ``simple and predictable''.
The implementation of AAEs in ASR systems has been proven to be more difficult than in image processing~\cite{cisse2017houdini}.
Some of them work irrespective of the architecture \cite{neekhara2019universal,vadillo2019universal,abdoli2019universal}.
However, these examples are crafted and tested using simplified architectures, either RNN or CNN.
They lack an attention mechanism, which is a relevant component of the framework used in our work.
Other works focus on making AAEs remain undetected by human subjects, e.g., by {psychoachustic hiding}~\cite{schonherr2018adversarial,qin2019imperceptible}.
Carlini et al.~\cite{carlini2016hidden} demonstrated how to extract AAEs for the CTC-based DeepSpeech architecture~\cite{Hannun2014DeepSS} by applying the FGSM to CTC loss.
Hu et al. gives a general overview over adversarial attacks on ASR systems and possible defense mechanisms in~\cite{hu2019adversarial}.
In it, they observe that by treating the features matrix of the audio input as the AAE seed, it is possible to generate AAE with algorithms developed in the image processing field.
However, this leads to the incapacity of the AAE to be transformed back to audio format, as the feature extraction of log-mel f-bank features is lossy.
Some have proposed ways to overcome this problem~\cite{Andronic2020}.
AAEs on the sequence-to-sequence attention-based LAS model~\cite{chan2016listen} by extending FGSM to attention are presented in~\cite{sun2019adversarial}.
%
%
In the same work, Sun et al. also propose adversarial regulation to improve model robustness by feeding back AAEs into the training loop.

\section{Audio Adversarial Example (AAE) Generation}
The following paragraphs describe the proposed algorithms to generate AAEs
(a) from two attention-based gradient methods, either using a static or a moving window adversarial loss;
(b) from a CTC-based FGSM, and
(c) combining both previous approaches in a joint CTC/attention approach.
In general, those methods apply the single-step FGSM~\cite{goodfellow2014explaining} on audio data and generate an additive adversarial noise $\bm{\delta}(\bm{x}_t)$ from a given audio feature sequence $\bm{X}=\bm{x}_{1:T}$, i.e.,
\begin{equation}
  \hat{\bm{x}}_t = \bm{x}_t + \bm{\delta}(\bm{x}_t),\hspace{0.02\textwidth} \forall t\in[1,T].
\end{equation}
We assume a \textit{whitebox} model, i.e., model parameters are known, to perform backpropagation through the neural network.
For any AAE algorithm, its reference sentence $y_{1:L}^*$ is derived from the network by decoding $\bm{x}_{1:T}$, instead of the ground truth sequence, to avoid label leaking~\cite{kurakin2016adversarial}.

\subsection{Attention-based Static Window AAEs}
For attention-based AAEs, the cross-entropy loss $\text{J}(\bm{X}, y_{l}; \bm{\theta} )$ w.r.t. $\bm{x}_{1:T}$ is extracted by iterating over {sequential token posteriors $p(y^*_{l}|y^*_{1:(l-1)})$} obtained from the attention decoder.
Sequence-to-sequence FGSM, as proposed in~\cite{sun2019adversarial}, then calculates $\bm{\delta}(\bm{x}_t)$  from the \emph{total} sequence as
\begin{align}\label{eq:seq-2-seq-fsgm}
    \bm{\delta}(\bm{x}_t) &= \epsilon \cdot \sgn(\nabla_{\bm{x}_t} \sum_{l = 1}^{L} J(\bm{X}, y_l^*; \bm{\theta} )), \quad l\in [1;L].
\end{align}
As opposed to this algorithm, our approach does not focus on the total token sequence, but only a portion of certain sequential steps.
This is motivated by the observation that attention-based decoding is auto-regressive;
interruption of the attention mechanism targeted at one single step in the sequence can change the corresponding portion of the transcription as well as throw off further decoding up to a certain degree.
A sum over all sequence parts as in Eq.~\ref{eq:seq-2-seq-fsgm} may dissipate localized adversarial noise. 
From this, the first attention-based method is derived that takes a single portion out of the output sequence.
We term this algorithm in the following as \emph{static window}  method.
Gradients in the sentence are summed up from the start token $ \gamma $ on to the following $ l_w $ tokens, such that
\begin{equation}\label{eq:window-static}
    \bm{\delta}_{\text{SW}}(\bm{x}_t) = \epsilon\cdot \sgn(\nabla_{\bm{x}_t} \sum_{l = \gamma}^{l_w} J(\bm{X}, y_l^*; \bm{\theta} )), \quad  l\in [1;L].
\end{equation}

\subsection{Attention-based Moving Window AAEs}
As observed from experiments with the static window, the effectiveness of the {static window} method strongly varies depending on segment position.
Adversarial loss from some segments has a higher impact than from others.
Some perturbations only impact local parts of the transcription.
Therefore, as an extension to the static window gradient derived from Eq.~\ref{eq:window-static}, multiple segments of the sequence can be selected to generate $\bm{\delta}_{MW}(\bm{x}_t)$.
We term this the {\emph{moving window}} method.
For this, gradients from a sliding window with a fixed length $l_w$ and {stride} $ \nu $ are accumulated to $  \nabla_{\text{MW}}(\bm{x}_t) $. 
The optimal values of length and stride are specific to each sentence.
Similar to the iterative FGSM based on momentum~\cite{dong2018boosting}, gradient normalization is applied in order to accumulate gradient directions.
\begin{align}
    \label{eq:window-moving}
    \nabla_{\text{MW}}(\bm{x}_t) &= \sum_{i = 0}^{\lceil L/\nu \rceil} \left( 
    \frac{\nabla_{\bm{x}_t}  \sum\limits_{l = i\cdot\nu}^{l_w} J(\bm{X}, y_l^*; \bm{\theta} ) }
    {||\nabla_{\bm{x}_t} \sum\limits_{l = i\cdot\nu}^{l_w} J(\bm{X}, y_l^*; \bm{\theta} )||_1}
    \right), \quad  l\in [1;L]\\
    \bm{\delta}_{MW}(\bm{x}_t) &= \epsilon\cdot \sgn( \nabla_{\text{MW}}(\bm{x}_t) )
\end{align}

\subsection{AAEs from Connectionist Temporal Classification}
From regular CTC loss $\Loss_{\text{CTC}}$ over the total reconstructed label sentence $\bm{y}^*$, the adversarial noise is derived as
\begin{align}
	\label{eq:ctc-fsgm}
    \bm{\delta}_{\text{CTC}}(\bm{x}_t) &= \epsilon\cdot \sgn(\nabla_{\bm{x}_t} \Loss_{\text{CTC}}(\bm{X}, \bm{y}^*; \bm{\theta} )).
\end{align}

\subsection{Hybrid CTC/Attention Adversarial Examples}
\begin{figure}[!htb]
  \centering
  \includegraphics[width=0.95\linewidth]{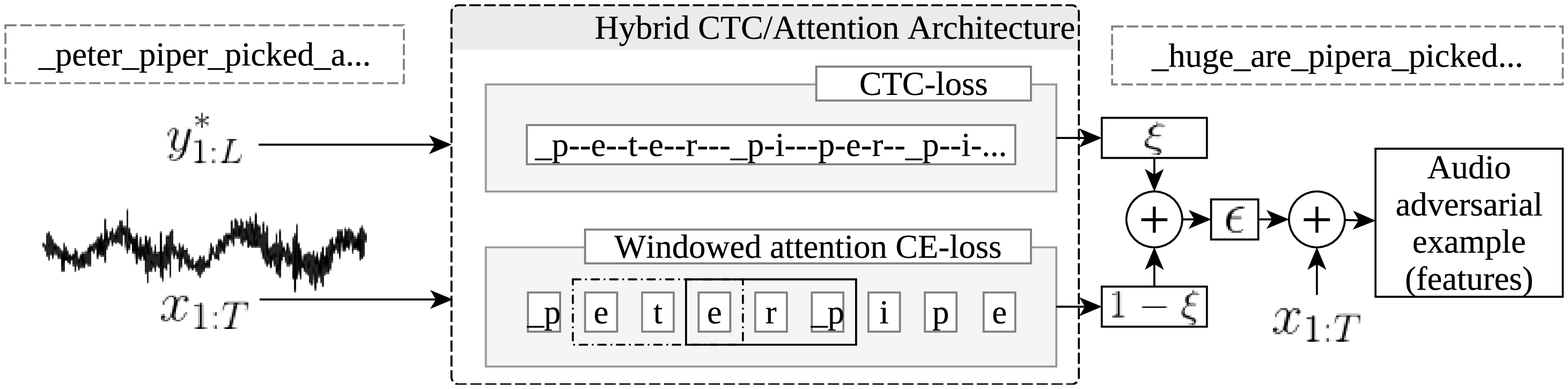}
  \caption{Generation of AAEs.
  The unmodified speech sentence $\bm{x}_{1:T}$ and the reference sentence $\bm{y}^*_{1:L}$ are given as input.
Then, using the hybrid CTC/attention model, the adversarial loss by the CTC-layer as well as the windowed attention sequence parts are calculated.
Those are then combined by the {weighting parameter $\xi$} and noise factor $\epsilon$ to obtain the adversarial example.
}
  \label{fig:advexgeneration}
\end{figure}

A multi-objective optimization function~\cite{lu2017multitask} is then applied to combine CTC and attention adversarial noise $\bm{\delta}_{\text{att}}$, that was either generated from $\bm{\delta}_{\text{SW}}$ or from $\bm{\delta}_{\text{MW}}$, by introducing the factor $\xi \in [0;1]$.
\begin{align}
	\label{eq:hybrid-aae}
    \hat{\bm{x}}_t = \bm{x}_t + (1-\xi)\cdot\bm{\delta}_{\text{att}}(\bm{x}_t) + \xi\cdot\bm{\delta}_{\text{CTC}}(\bm{x}_t), \hspace{0.02\textwidth} \forall t\in[1,T]
\end{align}
The full process to generate hybrid CTC/attention AAEs is shown in Fig.~\ref{fig:advexgeneration}.

\subsection{Adversarial Training}
Similar as data augmentation, Adversarial Training (AT) augments samples of a minibatch with adversarial noise $\bm{\delta}_{\text{AAE}}(x_t)$.
The samples for which we create the AAEs are chosen randomly with a probability of $p_a$, as proposed by Sun et al.~\cite{sun2019adversarial}.
Because of its successive backpropagation in a single step, this method is also termed \textit{adversarial regularization} and is applied not from the beginning of the neural network training but after the $N$th epoch.
Sun et al. additionally included a weighting factor $\alpha$ to distinct sequence components that we omit, i.e., set to $1$; instead, the gradient is calculated for the minibatch as a whole.
Furthermore, our AT algorithm also samples randomly the perturbation step size $\epsilon$ to avoid overfitting as originally described in~\cite{kurakin2016adversarial}.
The expanded gradient calculations for the {sequence-based training} is then written as
\begin{equation}\label{eq:adv_training_seq2seq}
\hat{J}(\bm{X}, y; \bm{\theta}) = \sum_i (J(\bm{X}, y_i; \bm{\theta}) + J(\hat{\bm{X}}, y_i; \bm{\theta})).
\end{equation}

\section{Evaluation}
Throughout the experiments, the hybrid CTC/attention architecture is used with an LSTM encoder with projection neurons in the encoder and location-aware attention mechanism, classifying from log-mel f-bank features~\cite{watanabe2017hybrid}.
As we evaluate model performance, and not human perception on AAEs, we limit our investigation to the feature space.
Evaluation is done on the TEDlium v2 \cite{rousseau2014enhancing} speech recognition task consisting of over $200$h of speech.
The baseline model we compare our results to is provided by the ESPnet toolkit~\cite{watanabe2018espnet}.
It has each four encoder and one an attention decoder layers with each $1024$ units per layer, and in total $50$m parameters.
We use the  BLSTM architecture for our experiments with each two layers in the encoder and the location-aware attention decoder;
the number of units in encoder, decoder and attention layers was reduced to 512 units~\cite{Chavez2020}.
This results in a model that has only one quarter in size compared to the baseline model, i.e., $14$m parameters.
For both models, $500$ unigram units serve as text tokens, as extracted from the corpus with SentencePiece~\cite{kudo2018subword}.
In all experiments, the reference token sequence $\bm{y^*}$ is previously decoded using the attention mechanism, as this is faster than hybrid decoding and also can be done without the RNNLM.
We also set $\epsilon = 0.3$, as done in AAE generation for attention-based ASR networks~\cite{sun2019adversarial}. 
Throughout the experiments, we use the decoded transcription $\bm{y^*}$ as reference, to avoid label leaking.
The dataset used in the evaluation, TEDlium v2, consists of recordings from presentations in front of an audience and therefore is already noisy and contains reverberations.
To better evaluate the impact of adversarial noise generated by our algorithms, two noise-free sample sentences are used for evaluation. 
Both sample sentences are created artificially using Text-to-Speech (TTS) toolkits so that they remain noise-free.

\subsection{Generation of AAEs: Two Case Studies}
The first noise-free sentence \emph{Peter} is generated from the TTS algorithm developed by Google named Tacotron 2~\cite{shen2018natural}.
It was generated using the pre-trained model by Google\footnote{https://google.github.io/tacotron/publications/tacotron2/index.html} and reads ``\emph{Peter Piper picked a peck of pickled peppers. How many pickled peppers did Peter Piper pick?}''
The second sentence \emph{Anie} was generated from the ESPNet TTS\footnote{The \texttt{ljspeech.tacotron2.v2} model.} and reads ``\emph{Anie gave Christina a present and it was beautiful.}''
We first test the CTC-based algorithm. 
The algorithm outputs for \emph{Peter} an AAE that has $41.3\%$ CER w.r.t. the ground-truth, whereas an error rate of $36.4\%$ for \emph{Anie}. 
For our experiments with the static window algorithm, we observe that it intrinsically runs the risk of changing only local tokens.
We take, for example, the sentence \textit{Anie} and set the parameter of $l_w = 3$ and $\gamma = 4$.
This gives us the following segment, as marked in bold font, out of the previously decoded sequence $\bm{y^*}$

\vspace{0.1cm}
\centerline{\emph{any gave ch{\textbf{ristin}}a a present and it was beautiful.}}
\vspace{0.1cm}
After we compute the AAE, the ASR system transcribes

\vspace{0.1cm}
\centerline{\emph{any game christian out priasant and it was beautiful}}
\vspace{0.1cm}

as the decoded sentence.
We obtain a sequence that strongly resembles the original where most of the words remain intact, while some of them change slightly.
Translated to CER and WER w.r.t the original sequence, we have 50.0 and 55.6 respectively.
We also test its hybrid version given $\xi = 0.5$, which is analogue to the decoding configuration of the baseline model.
It outputs a sequence with rates of $31.8\%$ CER, lower than its non-hybrid version. 
The moving window method overcomes this problem, as it calculates a non-localized AAE.
For example, a configuration with the parameters $\nu = 4 $ and $l_w = 2$ applied to \emph{Peter} generates the pattern

\vspace{0.1cm}
\centerline{\emph{\textbf{pe}ter pip\textbf{er p}icked a p\textbf{eck} of pickle\textbf{ pe}ppers.}}
\centerline{\emph{\textbf{ many p}ickle pe\textbf{pp}ers did pe\textbf{ter p}iper pa\textbf{ck}}}
\vspace{0.1cm}

for which we obtain the decoded sentence

\vspace{0.1cm}
\centerline{\emph{huter reperber picked a pick of piggle pebpers. }}
\centerline{\emph{how many tickle taper state plea piper pick.}}
\vspace{0.1cm}

This transcribed sentence then exhibits a CER of $54.3\%$ w.r.t the ground-truth.
The same parameter configuration applied in the hybrid version with $\xi = 0.5$ achieves error rates of $34.8\%$ CER.
Throughout the experiments, higher error rates were observed on the moving window than static window or CTC-based AAE generation.

\subsection{Evaluation of Adversarial Training}\label{subsec:at}
Throughout the experiments, we configured the {moving window method} with $\nu = 2$ and $l_w = 4 $ as arbitrary constant parameters,
motivated by the observation that those parameters performed well on both sentences \emph{Peter} and \emph{Ani}.
By inspection, this configuration is also suitable for sentences of the TEDlium v2 dataset.
Especially for its shorter utterances, a small window size and overlapping segments are effective.
Each model is trained for $10$ epochs, of which $N=5$ epochs are done in a regular fashion from regular training data;
then, the regular minibatch is augmented with its adversarial counterpart with a probability $p_a=0.05$.
Adversarial training examples are either attention-only, i.e. $\xi=0$, or hybrid, i.e. $\xi=0.5$.
Finally, the noise factor $\epsilon$ is sampled uniformly from a range of $[0.0;0.3]$ to cover a wide range of possible perturbations.
The trained model is compared with the baseline model as reported in~\cite{watanabe2018espnet}.
We use the moving window and its hybrid in the AT algorithm, because we hypothesize that both can benefit the training process of the hybrid model.
The RNNLM language model that we use is provided by the ESPnet toolkit~\cite{watanabe2018espnet};
it has $2$ layers with each $650$ units and its weight in decoding was set to $\beta=1.0$ in all experiments.
We did \emph{not} use data augmentation, such as SpecAugment, or language model rescoring;
both are known to improve ASR results, but we omit them for better comparability of the effects of adversarial training.
Results are collected by decoding four datasets:
(1) the regular test set of the TEDlium v2 corpus;
(2) AAEs from the test set, made with the {attention}-based {moving window} algorithm;
(3) the test set augmented with regular white noise at 30 dB SNR; and
(4) the test set with clearly noticeable 5 dB white noise.


\begin{table}[tb!]
	\centering
	\setlength{\tabcolsep}{4pt}
	\caption{Decoding results for all models.
	The first value in each cell corresponds to the CER and the second to the WER.
	The parameter $\lambda$ determines the weight of the CTC model during the decoding.
	Trained models with attention-only AAEs are marked with $\xi=0$; with hybrid AAEs with $\xi=0.5$.}
	\label{tab:decode_results_at} 
	\begin{tabular}{c c c c c c c c}
	    &  & & & \multicolumn{4}{c}{Dataset} \\
		\cmidrule(lr){5-8}
		\textbf{CER/WER}& $\xi$ & $\lambda$ & LM & test & \begin{tabular}{@{}c@{}}test \\ AAE\end{tabular} & \begin{tabular}{@{}c@{}}noise \\ 30dB \end{tabular} & \begin{tabular}{@{}c@{}}noise \\ 5dB \end{tabular} \\ 
		\cmidrule(lr){2-8} 
		\multirow{3}{*}[-1pt]{baseline~\cite{watanabe2018espnet}} & - & 0.0 & \textbf{-} & 20.7/22.8 & 90.7/89.1          & 23.6/25.8 & 78.8/78.8  \\ 
		    & - & 0.5 & \textbf{-} & 15.7/18.6 & 86.1/89.9  & 18.1/21.3 & 66.1/68.3  \\ 
		    & -&  0.5 & \checkmark & 16.3/18.3 & \textbf{98.5/92.2} & 19.2/20.8 & 73.2/72.7\\
		\midrule 
		\multirow{3}{*}[-1pt]{\parbox{2.5cm}{adv. trained with att.-only AAE}} & 0.0 & 0.0 & \textbf{-} & 17.7/19.6 & 63.6/63.3 & 21.0/22.8  & 74.7/74.4 \\ 
	    	& 0.0 & 0.5 & \textbf{-} & 14.3/16.9 & \textbf{53.5/56.8} & 16.5/18.9  & 62.6/65.0  \\ 
	    	& 0.0 & 0.5 & \checkmark & 15.1/16.9 & 60.3/58.3 & 17.5/18.9  & 69.0/68.0\\
	    \midrule
	    \multirow{3}{*}[-1pt]{\parbox{2.5cm}{adv. trained with hybrid AAE}} & 0.5 & 0.0 & \textbf{-} & 17.9/19.8 & 65.2/65.0 & 20.4/22.3 & 74.9/75.0 \\ 
	    	& 0.5 & 0.5 & \textbf{-} & \textbf{14.0/16.5} & \textbf{54.8/58.6} & \textbf{16.2/18.7} & \textbf{63.5/65.8} \\ 
	    	& 0.5 & 0.5 & \checkmark & {14.8/16.6} & 61.8/59.9 & 17.0/18.5 & 70.0/69.2\\
	\bottomrule	
	\end{tabular}
\end{table}


\paragraph{General trend.}
Some general trends during evaluation are manifested in Tab.~\ref{tab:decode_results_at}.
Comparing decoding performances between the regular test set and the AAE test set, all models perform worse.
In other words, the moving window technique used for creating the AAEs performs well against different model configurations.
Setting the CTC weight lowers error rates in general.
The higher error rates in combination with a LM are explained by the relatively high weight $\beta=1.0$.
Rescoring leads to improved performance, however, listed results are more comparable to each other when set to a constant in all decoding runs.

\paragraph{Successful AAEs.}
Notably, the baseline model performed worst on this dataset with almost $100\%$ error rates, even worse when decoding noisy data.
This manifests in wrong transcriptions of around the same length as the ground truth, with on average $90\%$ substitution errors but only $20\%$ insertion or deletion errors.
Word loops or dropped sentence parts were observed only rarely, two architectural behaviors when the attention decoder looses its alignment.
We report CERs as well as WERs, as a relative mismatch between those indicates certain error patterns for CTC and attention decoding~\cite{kurzinger2019exploring};
however, the ratio of CER to WER of transcribed sentences was observed to stay roughly at the same levels in the experiments with the AAE test set~\cite{kurzinger2019exploring}

\paragraph{Adv. trained models are more robust.}
Both models obtained from adversarial training perform better in general, especially in the presence of adversarial noise, than the baseline model;
the model trained with hybrid AAEs achieves a WER of $16.5\%$ on the test set, a performance of absolute $1.8\%$ over the baseline model.
At the same time, the robustness on regular noise and specially on adversarial noise was improved.
For the latter we have an improvement of $ 24-33\% $ absolute WER.
The most notable difference is in decoding in combination with CTC and LM, where the regularly trained model had a WER of $92.2\%$, while the corresponding adv. trained model had roughly $60\%$ WER. 
The att.-only adv. trained model with $\xi=0$ seems to be slightly more robust.
On the one hand that might be a side effect from the the AAEs that are generated in an attention-only manner;
on the other hand, this model also slightly performed better on regular noise.

\section{Conclusion}
In this work, we demonstrated audio adversarial examples against hybrid attention/CTC speech recognition networks.
The first method we introduced was to select a \emph{static window} over a selected segment of the attention-decoded output sequence to calculate the adversarial example.
This method was then extended to a \emph{moving window} that slides over the output sequence to better distribute perturbations over the transcription.
In a third step, we applied the fast gradient sign method to CTC-network.

AAEs constructed with this method induced on a regular speech recognition model a word error rate of up to $90\%$.
In a second step, we employed these for adversarial training a hybrid CTC/attention ASR network.
This process improved its robustness against audio adversarial examples, with $55\%$ WER, and also slightly against regular white noise.
Most notably, the speech recognition performance on regular data was improved by absolute $1.8\%$ from $18.3\%$ compared to baseline results.

\bibliographystyle{splncs04}
\bibliography{paper}

\begin{thebibliography}{10}
\providecommand{\url}[1]{\texttt{#1}}
\providecommand{\urlprefix}{URL }
\providecommand{\doi}[1]{https://doi.org/#1}

\bibitem{abdoli2019universal}
Abdoli, S., Hafemann, L.G., Rony, J., Ayed, I.B., Cardinal, P., Koerich, A.L.:
  Universal adversarial audio perturbations. ArXiv  \textbf{abs/1908.03173}
  (2019)

\bibitem{Andronic2020}
Andronic, I.: {MP3 Compression as a Means to Improve Robustness against
  Adversarial Noise Targeting Attention-based End-to-End Speech Recognition}.
  Master's thesis, Technical University of Munich, Germany (2020)

\bibitem{BahdanauEtAl14}
Bahdanau, D., Cho, K., Bengio, Y.: Neural machine translation by jointly
  learning to align and translate. arXiv preprint arXiv:1409.0473  (2014)

\bibitem{carlini2016hidden}
Carlini, N., Mishra, P., Vaidya, T., Zhang, Y., Sherr, M., Shields, C., Wagner,
  D., Zhou, W.: Hidden voice commands. In: 25th $\{$USENIX$\}$ Security
  Symposium ($\{$USENIX$\}$ Security 16). pp. 513--530 (2016)

\bibitem{carlini2018audio}
Carlini, N., Wagner, D.: Audio adversarial examples: Targeted attacks on
  speech-to-text. In: 2018 IEEE Security and Privacy Workshops (SPW). pp.~1--7.
  IEEE (2018)

\bibitem{chan2016listen}
Chan, W., Jaitly, N., Le, Q., Vinyals, O.: Listen, attend and spell: A neural
  network for large vocabulary conversational speech recognition. In: 2016 IEEE
  International Conference on Acoustics, Speech and Signal Processing (ICASSP).
  pp. 4960--4964. IEEE (2016)

\bibitem{Chavez2020}
Chavez~Rosas, E.R.: {Improving Robustness of Sequence-to-sequence Automatic
  Speech Recognition by Means of Adversarial Training}. Master's thesis,
  Technical University of Munich, Germany (2020)

\bibitem{ChorowskiEtAl15}
Chorowski, J.K., Bahdanau, D., Serdyuk, D., Cho, K., Bengio, Y.:
  Attention-based models for speech recognition. neural information processing
  systems pp. 577--585 (2015)

\bibitem{cisse2017houdini}
Cisse, M., Adi, Y., Neverova, N., Keshet, J.: Houdini: Fooling deep structured
  prediction models. ArXiv  \textbf{abs/1707.05373} (2017)

\bibitem{dong2018boosting}
Dong, Y., Liao, F., Pang, T., Su, H., Zhu, J., Hu, X., Li, J.: Boosting
  adversarial attacks with momentum. In: Proceedings of the IEEE conference on
  computer vision and pattern recognition. pp. 9185--9193 (2018)

\bibitem{goodfellow2014explaining}
Goodfellow, I.J., Shlens, J., Szegedy, C.: Explaining and harnessing
  adversarial examples. CoRR  \textbf{abs/1412.6572} (2014)

\bibitem{graves2006connectionist}
Graves, A., Fern{\'a}ndez, S., Gomez, F., Schmidhuber, J.: Connectionist
  temporal classification: labelling unsegmented sequence data with recurrent
  neural networks. In: Proceedings of the 23rd international conference on
  Machine learning. pp. 369--376. ACM (2006)

\bibitem{GulcehreEtAl15}
Gulcehre, C., Firat, O., Xu, K., Cho, K., Barrault, L., Lin, H.C., Bougares,
  F., Schwenk, H., Bengio, Y.: On using monolingual corpora in neural machine
  translation. arXiv preprint arXiv:1503.03535  (2015)

\bibitem{Hannun2014DeepSS}
Hannun, A.Y., Case, C., Casper, J., Catanzaro, B., Diamos, G., Elsen, E.,
  Prenger, R., Satheesh, S., Sengupta, S., Coates, A., Ng, A.Y.: Deep speech:
  Scaling up end-to-end speech recognition. ArXiv  \textbf{abs/1412.5567}
  (2014)

\bibitem{hu2019adversarial}
Hu, S., Shang, X., Qin, Z., Li, M., Wang, Q., Wang, C.: Adversarial examples
  for automatic speech recognition: Attacks and countermeasures. IEEE
  Communications Magazine  \textbf{57}(10),  120--126 (2019)

\bibitem{kudo2018subword}
Kudo, T.: Subword regularization: Improving neural network translation models
  with multiple subword candidates. ArXiv  \textbf{abs/1804.10959} (2018),
  dOI:10.18653/v1/P18-1007

\bibitem{kurakin2016adversarialphysical}
Kurakin, A., Goodfellow, I., Bengio, S.: Adversarial examples in the physical
  world. CoRR  \textbf{abs/1607.02533} (2016),
  \url{http://arxiv.org/abs/1607.02533}

\bibitem{kurakin2016adversarial}
Kurakin, A., Goodfellow, I., Bengio, S.: Adversarial machine learning at scale.
  CoRR  \textbf{abs/1611.01236} (2016), \url{http://arxiv.org/abs/1611.01236}

\bibitem{kurzinger2019exploring}
K{\"u}rzinger, L., Watzel, T., Li, L., Baumgartner, R., Rigoll, G.: Exploring
  hybrid ctc/attention end-to-end speech recognition with gaussian processes.
  In: International Conference on Speech and Computer. pp. 258--269. Springer
  (2019)

\bibitem{lu2017multitask}
Lu, L., Kong, L., Dyer, C., Smith, N.A.: Multitask learning with ctc and
  segmental crf for speech recognition  (2017). \doi{10/gf3hs6}

\bibitem{neekhara2019universal}
Neekhara, P., Hussain, S., Pandey, P., Dubnov, S., McAuley, J., Koushanfar, F.:
  Universal adversarial perturbations for speech recognition systems. ArXiv
  \textbf{abs/1905.03828} (2019), dOI: 10.21437/interspeech.2019-1353

\bibitem{qin2019imperceptible}
Qin, Y., Carlini, N., Goodfellow, I., Cottrell, G., Raffel, C.: Imperceptible,
  robust, and targeted adversarial examples for automatic speech recognition.
  ArXiv  \textbf{abs/1903.10346} (2019)

\bibitem{rousseau2014enhancing}
Rousseau, A., Del{\'e}glise, P., Esteve, Y.: Enhancing the ted-lium corpus with
  selected data for language modeling and more ted talks. In: LREC. pp.
  3935--3939 (2014)

\bibitem{schonherr2018adversarial}
Sch{\"o}nherr, L., Kohls, K., Zeiler, S., Holz, T., Kolossa, D.: Adversarial
  attacks against automatic speech recognition systems via psychoacoustic
  hiding. ArXiv  \textbf{abs/1808.05665} (2018), dOI:10.14722/ndss.2019.23288

\bibitem{shen2018natural}
Shen, J., Pang, R., Weiss, R.J., Schuster, M., Jaitly, N., Yang, Z., Chen, Z.,
  Zhang, Y., Wang, Y., Skerrv-Ryan, R., et~al.: Natural tts synthesis by
  conditioning wavenet on mel spectrogram predictions. In: 2018 IEEE
  International Conference on Acoustics, Speech and Signal Processing (ICASSP).
  pp. 4779--4783. IEEE (2018)

\bibitem{sun2019adversarial}
Sun, S., Guo, P., Xie, L., Hwang, M.Y.: Adversarial regularization for
  attention based end-to-end robust speech recognition. IEEE/ACM Transactions
  on Audio, Speech, and Language Processing  \textbf{27}(11),  1826--1838
  (2019)

\bibitem{szegedy2013intriguing}
Szegedy, C., Zaremba, W., Sutskever, I., Bruna, J., Erhan, D., Goodfellow, I.,
  Fergus, R.: Intriguing properties of neural networks. CoRR
  \textbf{abs/1312.6199} (2013)

\bibitem{vadillo2019universal}
Vadillo, J., Santana, R.: Universal adversarial examples in speech command
  classification. ArXiv  \textbf{abs/1911.10182} (2019)

\bibitem{watanabe2018espnet}
Watanabe, S., Hori, T., Karita, S., Hayashi, T., Nishitoba, J., Unno, Y.,
  {Enrique Yalta Soplin}, N., Heymann, J., Wiesner, M., Chen, N.,
  Renduchintala, A., Ochiai, T.: Espnet: End-to-end speech processing toolkit.
  In: Interspeech. pp. 2207--2211 (2018),
  \url{\url{http://dx.doi.org/10.21437/Interspeech.2018-1456}}, dOI:
  10.21437/Interspeech.2018-1456

\bibitem{watanabe2017hybrid}
Watanabe, S., Hori, T., Kim, S., Hershey, J.R., Hayashi, T.: Hybrid
  ctc/attention architecture for end-to-end speech recognition. IEEE Journal of
  Selected Topics in Signal Processing  \textbf{11}(8),  1240--1253 (2017)

\end{thebibliography}
\end{document}